\documentclass[preprint, 12pt]{elsarticle}

\usepackage{amssymb}
\usepackage{amsmath}

\usepackage[most]{tcolorbox}
\usepackage{dirtytalk}
\usepackage{kotex}
\usepackage{subcaption}
\usepackage{tabularray}
\usepackage{longtable}
\usepackage{booktabs}

\DeclareMathOperator*{\argmax}{arg\,max}

\begin{document}

\begin{frontmatter}

\title{Wallets' Explorations Across Non-Fungible Token Collections}

\author[pp]{Seonbin Jo}
\affiliation[pp]{organization={Department of Physics, Pohang University of Science and Technology},
            addressline={77 Cheongam-Ro, Nam-Gu}, 
            city={Pohang},
            state={Gyeongbuk},
            country={Republic of Korea}}

\author[pp,pime]{Woo-Sung Jung}
\affiliation[pime]{organization={Department of Industrial and Management Engineering, Pohang University of Science and Technology},
            addressline={77 Cheongam-Ro, Nam-Gu}, 
            city={Pohang},
            state={Gyeongbuk},
            country={Republic of Korea}}

\author[elon]{Hyunuk Kim\corref{cor1}}
\ead{hkim6@elon.edu}
\cortext[cor1]{Corresponding Author}
\affiliation[elon]{organization={Department of Management and Entrepreneurship, Love School of Business, Elon University},
            addressline={2075 Campus Box}, 
            city={Elon}, 
            state={NC},
            country={USA}}

\begin{abstract}
Non-fungible tokens (NFTs), which are immutable and transferable tokens on blockchain networks, have been used to certify the ownership of digital images often grouped in collections. Depending on individual interests, wallets explore and purchase NFTs in one or more image collections. Among many potential factors of shaping purchase trajectories, this paper specifically examines how visual similarities between collections affect wallets' explorations. Our model shows that wallets' explorations are not random but tend to favor collections having similar visual features to their previous purchases. The model also predicts the extent to which the next collection is close to the most recent collection of purchases with respect to visual features. These results are expected to enhance and support recommendation systems for the NFT market.
\end{abstract}


\end{frontmatter}

\section{Introduction}
\label{Introduction}

Non-fungible tokens (NFTs) are immutable and traceable objects on blockchain networks so that can be used to prove the ownership of digital assets, including artwork, photo, audio, and video. Nowadays, most NFTs are in collections sharing similar concepts and traits, such as hat, glass, and background color. For example, a famous collection `Bored Ape Yacht Club' is made of NFTs of cartoon apes. Artists creating collections sell NFTs to wallets, and these wallets sell NFTs to other wallets through marketplaces referred to as the NFT market in this paper. 

The prices of NFTs are associated with visual features (e.g., pixel, color)~\citep*{nadini_mapping_2021, fridgen_pricing_2023}. The extent to which traits and their combinations are rare affects NFTs' market values~\citep{mekacher_heterogeneous_2022}. New NFT collections inspired by the visual design of existing collections tend to have similar financial performance to the existing collections~\citep{la_cava_visually_2023}. Considering that the price of an NFT is determined by its demand, visual features may also influence demand and individual wallet's purchase decisions, specifically whether to purchase NFTs having similar or different visual features. 

Wallets face diverse options at the moments of purchase. The uncertainty of choices leads different purchase trajectories by wallet. Some wallets possess only one NFT collection, while others purchase various collections. These two cases can be mapped to `Exploitation' and `Exploration', respectively, where exploitation is a behavior maximizing the current utility by re-using known options and exploration is a behavior taking risk for higher returns by searching uncertain environments~\citep{march_exploration_1991, cohen_should_2007}.

Decisions whether to choose exploration or exploitation are made based on environmental, social, individual factors~\citep{mehlhorn_unpacking_2015}. The belief constructed from the previous environment does not hold in the new environment and in turn would change the decisions~\citep{posen_chasing_2012}. Competition is another environmental factor. If competitions exist, the risk of exploration increases because there is a possibility that competitors already exploited the environment. Under competition, expected returns would decrease and lead to a tendency leaning toward exploitation~\citep{phillips_rivals_2014}. The risk of exploration can decrease if social information is available in decision-making~\citep{valone_group_1989}. If a person identifies the major decision in a system through social information, this person is more likely to follow the major decision~\citep{lee_exploration_2003}. It is because of network effects that increasing population which chooses the major decision is expected to receive higher returns compared to others choosing non-major decisions. Prior experience also affects decisions in the future. When customers of an online food delivery service explore new cuisines, they tend to choose cuisines similar to the foods they ordered~\citep{schulz_structured_2019}. 

In this paper, we focus on how wallets exploit an NFT collection and explore new collections based on a visual similarity. Environmental and social factors are surely important in the NFT market, but they are difficult to measure as the NFT market has evolved fast and does not reveal personal identities and their connections. Therefore, we set the research questions as:

\begin{enumerate}
    \item[\textbf{RQ1}.] How do exploration and exploitation occur in the NFT market? \\
    \item[\textbf{RQ2}.] Do wallets explore new NFTs in collections having visual features similar to their previous collections of purchases?
\end{enumerate}

Researchers found that behavioral trajectories in nature and social systems consist of many short movements and a few long jumps to distant areas with low probabilities. Honey bees~\citep{reynolds_displaced_2007}, desert ants~\citep{reynolds_optimal_2008}, sharks~\citep{sims_encounter_2006}, dears~\citep{focardi_adaptive_2009}, and elephants~\citep{dai_short-duration_2007} search resources by following this pattern. Humans are not an exception. They make long non-local jumps occasionally in physical spaces~\citep{brockmann_scaling_2006, brown_levy_2007, gonzalez_understanding_2008} and virtual spaces~\citep{garg_efficient_2021}.

L\'evy flight~\citep{mandelbrot1982fractal} is a random walk model explaining this pattern with a probability distribution $d^{-\alpha}$ where $d$ is a distance between two points in a trajectory and $\alpha$ is a positive value between 1 and 3 in many real-world systems~\citep{garg_efficient_2021, zaburdaev_levy_2015}. This distribution generates frequent short movements and rare long jumps because of its fat tail. Borrowing the functional form of l\'evy flight, we aim to examine whether wallets explore the NFT market with the same pattern. If estimated $\alpha$ values for wallets are also between 1 and 3, we would argue that wallet owners' behave similarly with agents in many real-world systems. We used (1 - a visual similarity) as $d$ (see details in Section~\ref{DataMethods}).

\begin{enumerate}
    \item[\textbf{RQ3}.] Do wallets explore NFT collections similarly to agents in physical systems with the pattern of frequent short but rare long movements?
\end{enumerate}

\section{Data and Methods}
\label{DataMethods}

\subsection{NFT transfer data}
It is challenging to identify individual buyers for many physical artworks. For example, Picasso’s La Dormeuse was sold for 57.8 million US dollars, but we don't know who owns it~\cite{gleadell_phillips_2018}. In contrast, the NFT market is governed by smart contracts which record transactions in blockchain networks, so we can examine when a new NFT was minted to a wallet and which wallets purchased NFTs. Details of transactions such as fees and asset values are also available on blockchain networks. This property allows us to reconstruct each wallet's purchase trajectory even though we can't connect wallets with real-world personal identities.

\begin{figure}[h]
    \centering
    \includegraphics[width=\textwidth]{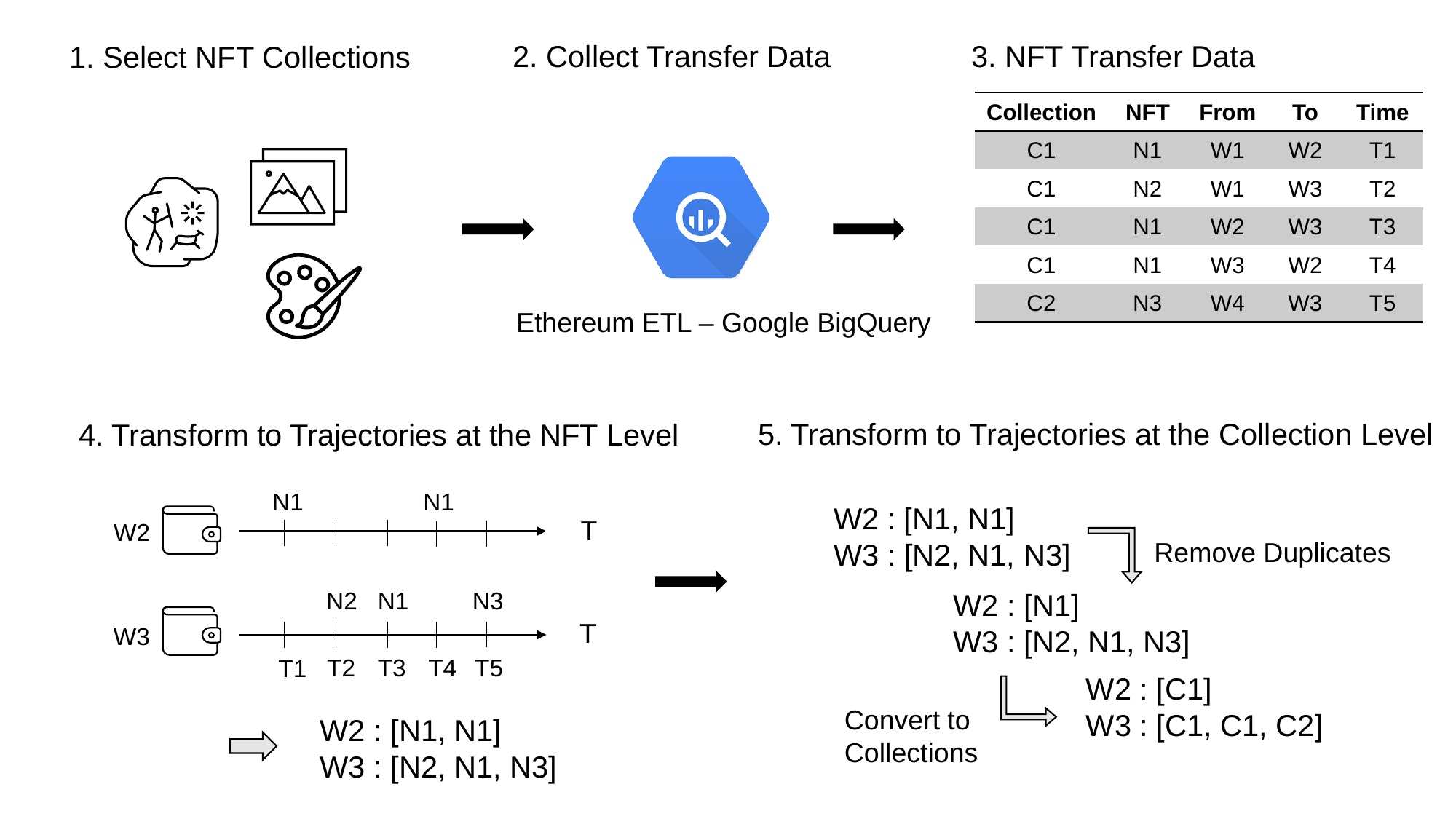}
    \caption{A schematic diagram of data collection and cleaning. First, we selected 198 NFT collections and gathered transfer data from the Ethereum ETL project in Google BigQuery. After collecting transfer data, we grouped rows by buyer (`To' columns in the table) and transformed them into purchase trajectories at the NFT level. Then, the purchase trajectories at the NFT level is converted to the trajectories at the collection level after removing duplicated NFT purchases.}
    \label{fig1}
\end{figure}

We specifically focused on the top 200 NFT collections with respect to volume (i.e., the total monetary value of transfers) and the top 200 NFT collections with respect to sales (i.e., the total number of transfers) in the Etherium network, according to the OpenSea ranking statistics on July 31, 2023~\cite{opensea_top_nodate}. We combined them into one list of NFT collections and excluded some NFT collections based on the following criteria (Figure~\ref{fig1}--Step 1). First, we removed collections that consist of nearly identical images. For instance, `Opepen Edition', `Checks - VV Edition', and `Damien Hirst - The Currency' that have simple features, such as random dots with only color variations, were excluded. Second, we removed NFT collections that have factors beyond image itself. For example, NFT collections about soccer players incorporate players' game performance in price so that were excluded not to confound our analysis. These steps yield a list of 198 NFT collections (See SI for the full list of NFT collections used in this paper). 

Two different token standards -- ERC-20 and ERC-721 -- are used for the selected collections. ERC-20 generates a new currency unit with its own standards in the network. Tokens associated with a system built on ERC-20 are interchangeable. In an ERC-20 system, 10 tokens of a wallet are worth as same as 10 tokens of another wallet. However, a token in an ERC-20 system can't be traded with a token in another ERC-20 system. Regarding our research, CryptoPunk is the only collection using ERC-20. Each digital artwork in CryptoPunk acts as an independent currency unit with a maximum supply of 1. This characteristic makes artworks not replicated and not interchangeable as they are different ERC-20 tokens. On the other hand, tokens with ERC-721 are not fungible and exist uniquely. Therefore, they represent the ownership of digital artworks. Most NFT collections are hosted with the ERC-721 standard.

We leveraged the Ethereum ETL project built upon Google BigQuery~\cite{noauthor_blockchain-etlethereum-etl_nodate} to collect NFT transactions associated with ERC-20 and ERC-721 (Figure~\ref{fig1}--Step 2). The raw data were then refined further to have direct transfers between individual wallets. This dataset includes information about sellers (from) and buyers (to) with timestamps for NFT transfers (Figure~\ref{fig1}--Step 3). From this dataset, we generated every individual's purchase trajectory by grouping transfers by buyer in the ascending order of time (Figure~\ref{fig1}--Step 4).

\subsection{Constructing purchase trajectories}

The main goal of this paper is to understand exploring behaviors in the NFT market, so we focused on unique NFTs in purchase trajectories (Figure~\ref{fig1}--Step 5). For example, the trajectory of Wallet 2 (W2) in Figure~\ref{fig1} is $[N1, N1]$, which means a re-purchase of $N1$, and we reduced the trajectory to $[N1]$. We also converted trajectories at the NFT level to those at the collection level (Figure~\ref{fig1}--Step 5). The trajectory of Wallet 3 (W3) in Figure~\ref{fig1} becomes $[C1, C1, C2]$. In our study, $[C1, C1]$ is an intra-collection jump corresponding exploitation and $[C1, C2]$ is an inter-collection jump corresponding exploration. 

Finally, we filtered out wallets of which collection-level trajectory having less than 10 or more than 100 collections. It is because that the trajectories having less than 10 collections are not enough for analysis and the trajectories having more than 100 collections are more likely to be made by bots, contract accounts, and marketplaces, which are beyond our scope. 159,689 distinct wallets remain for further analyses. 

\subsection{NFT and collection embeddings}

Image embedding is a technique projecting an image to a low-dimensional vector while preserving its features with a little loss. Similar images are positioned close on a vector space. We collected NFT images through the links provided by the OpenSea API and used a pre-trained ResNet18~\cite{he_deep_2016}, a deep neural network specialized for image classification task, to calculate NFT embeddings in 512 dimensions. 

NFT embeddings allow us to characterize collections as well. We obtained the centroid of NFT embeddings of a collection and define it as \textit{collection embedding}. An advantage of having collection embeddings is that we are able to quantify similarities between collections. In this paper, we used the cosine similarity between collection embeddings as a similarity measure. Suppose a wallet purchased an NFT in Collection A and then another in Collection B. This wallet's purchasing history can be described as a jump from A's embedding to B's embedding. If these collections are similar with respect to visual features, the jump distance from A to B is relatively short. However, if this wallet purchased an NFT in Collection C, which is significantly different from A and B, the jump distance from B to C is longer than the distance from A to B.

We validated the embedding space by randomly choosing three collections and comparing each collection with its five closest collections with respect to the cosine distance (1-cosine similarity; Table~\ref{tab:my_label}). `3Landers' has pastel colors, and its NFTs have a round cartoon design. `Dippies' is the most closest collection on the NFT embedding space to `3Landers' and shares similar colors and a round cartoon design. `Chain Runners' is composed with pixel arts, and its five closest collections have similar pixel designs. The closest neighbor of the final example `NotOkayBears' is `Okay Bears Yacht Club'. As the collection name implies, they are similar in many aspects, including character design. Other collections close to `NotOkayBears' on the embedding space also share a similar design with a famous NFT collection `Bored Ape Yacht Club'~\cite{la_cava_visually_2023}. These observations validate our NFT embedding space. 

\begin{table}[h]
    \small
    \centering
    \begin{tblr}{
    width=\textwidth,
    colspec = {|Q[c,m]|Q[c,m]|Q[c,m]|Q[c,m]|},
    cells   = {font = \fontsize{9pt}{12pt}\selectfont},
    hline{1,8} = {1pt,solid},
    hline{2,7} = {2pt,solid},
    hline{3-6} = {1pt,solid},
    }
         Sampled Collection & 3Landers     &  Chain Runners & NotOkayBears\\ 
         
         1st Closest & {inBetweeners \\ (0.0475)} & {Bears Deluxe \\ (0.0445)} &  {Okay Bears Yacht Club \\ (0.0166)}\\ 
         2nd Closest & {Dippies  \\ (0.0482)} & {Nakamigos \\ (0.0527)} &  {Fat Rat Mafia \\ (0.0604)}\\ 
         3rd Closest & {Deadfellaz \\ (0.0510)} & {Murakami.Flowers \\ (0.0538)} &  {Wicked Ape Bone Club \\ (0.0737)}\\ 
         4th Closest & {The Long-Lost \\ (0.0579)} & {CyberKongz \\ (0.0595)} &  {Bored Ape Yacht Club \\ (0.0737)}\\ 
         5th Closest & {DentedFeelsNFT \\ (0.0584)} & {Genuine Undead \\ (0.0705)} &  {Galaxy Fight Club \\ (0.0748)}\\ 
         Similar properties & {Color \& \\ Round cartoon design} & {Pixel arts} &  {Design similar to \\ Bored Ape Yacht Club}\\
    \end{tblr}
    \caption{Three sampled collections and their five closest collections on the embedding space. Values in parentheses are distances from the sampled collection. Each set of the five collections has a similar design and concepts. For example, the set associated with `3lenders' has a similar round cartoon design. The selected neighbors of `Chain Runners' are pixel arts, and those of `NotOkayBears' are similar to `Bored Ape Yacht Club'.}
    \label{tab:my_label}
\end{table}

\section{Modeling}

Wallets would have distinct purchase trajectories depending on their owners' preferences and personal characteristics. We developed three models generating synthetic purchase trajectories and compared them with the data to explain potential factors shaping the real purchase trajectories. The following notations are used for explaining our models. Each wallet has a trajectory $T$, which is a sequence of collections of NFTs that its owner purchased. $T_{i}$ is the trajectory of a wallet $i$, and $t_{i,j}$ is the $j$th collection in $T_{i}$. Trajectories have different length and starting collection, so for a wallet $i$, we fix both $|T_{i}|$ and $t_{i,1}$ in simulations to compare synthetic trajectories with the data fairly.

\begin{figure}[!]
    \centering
    \includegraphics[width=\textwidth]{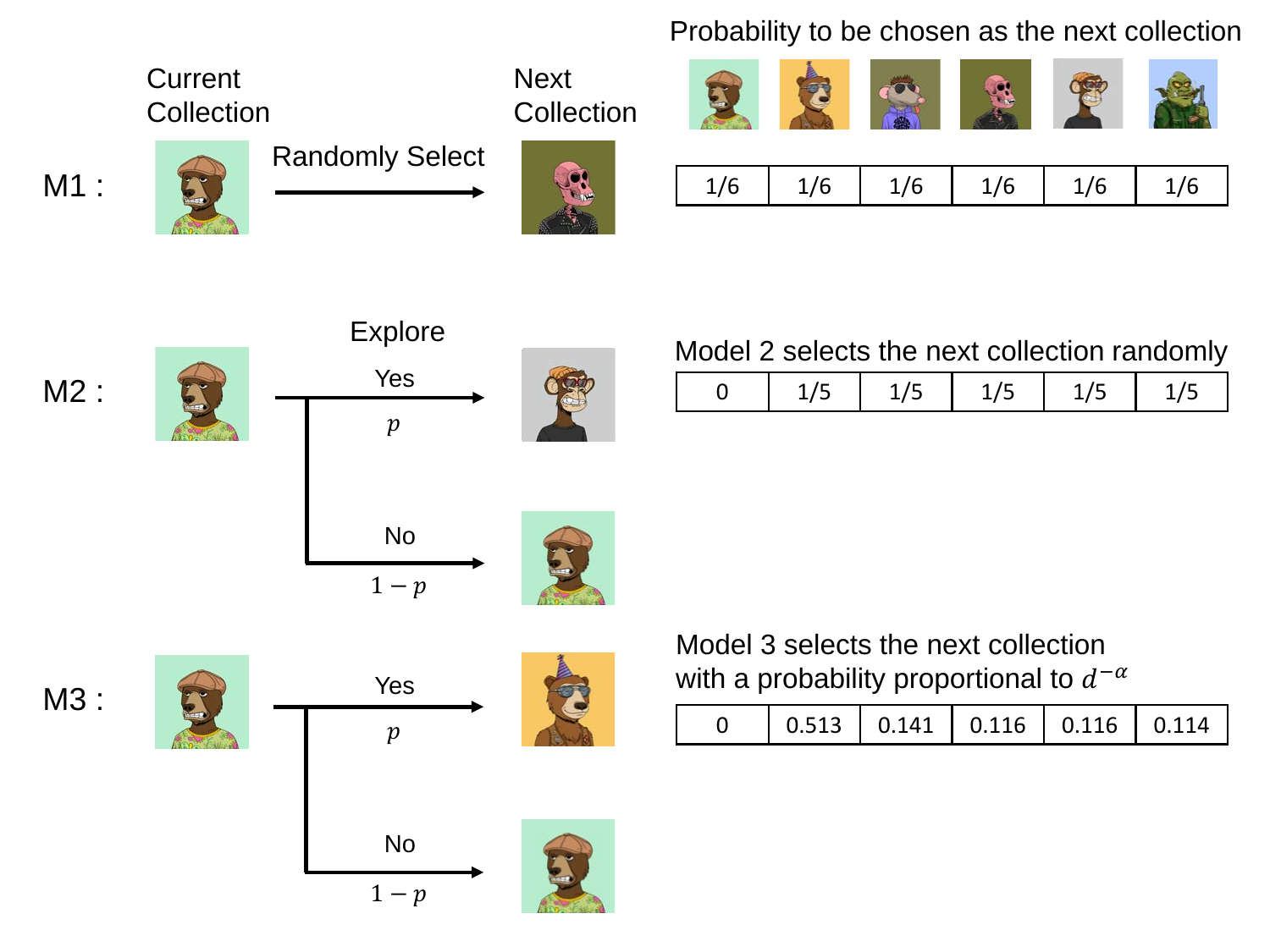}
    \caption{A schematic diagram of the models. The figure of NFTs are from `NotOkayBears'. Tables in the figure show probabilities that a collection is chosen as the next collection. Model 1 (M1) selects the next collection randomly. This model is the baseline for comparison with other models. Model 2 (M2) introduces an exploration parameter $p$, which decides the frequency of exploration. In this model, an agent explores a new collection with a probability $p$. In case of exploration, the next collection is selected randomly. Model 3 (M3) is similar to Model 2 but the next collection is sampled from a function following L\'evy flight with an exponent $\alpha$.
    }
    \label{model_fig}
\end{figure}

\subsection{Model 1}
The first model is a completely random model. For each wallet, after fixing the starting collection, we sampled collections randomly from the list of all collections $C$ with equal probabilities. The probability that a collection is chosen as the next collection is $1/|C|$. This model is simple but allows us to understand the NFT collection embedding space with the step length distribution, a distribution for jump distances within trajectories. For example, if a wallet has a sequence $[C_1, C_1, C_2, C_3]$, then a set of step lengths for this sequence becomes $\{d(C_1, C_1), d(C_1, C_2), d(C_2, C_3)\}$ where $d(X, Y)$ is the cosine distance between the collection embedding vectors of $X$ and $Y$. Note that $d(X, X)$ is zero. Since this model generates random sequences, the aggregated step length distribution from all wallets $f_{null}(d)$ represents probabilities to sample collections reachable at a distance $d$ (Figure~\ref{figure2}). This distribution is not biased by other external factors and represents the inherent topology of the embedding space.

\subsection{Model 2}
However, wallet owners' would not purchase NFTs randomly. They are likely to have different preferences on exploring and exploiting collections. To measure the preferences, for a wallet $i$, we calculated $p_i$ which is the proportion of choosing a different collection from the most recent purchase. For example, a wallet $i$ of which purchase trajectory is $\{C_1, C_1, C_2, C_3\}$ changes collections twice, $C_1 \rightarrow C_2$ and $C_2 \rightarrow C_3$, while the total number of changes is three. Then, $p_i$ becomes $2/3$. We assume that $p_i$ estimated from the data represents $i$'s unique characteristics so fix the values in modeling. 

The second model adds the effect of $p_i$ to the first model. It generates a sequence as follows. A wallet owner $i$ decides whether to buy an NFT in the same collection of the previous purchase or not. $i$ explores a new collection with a probability $p_i$ or exploits the collection of the most recent purchase with a probability $1-p_i$. In case of exploration, a collection is randomly chosen as the next collection with a probability $1/(|C|-1)$. Comparing the first and the second models, we can better understand the impact of $p_i$.

\subsection{Model 3}
Lastly, we incorporated personal characteristics about selecting a new collection. Some wallet owners tend to choose a collection similar to the previous collection, while others tend to choose a collection different from the previous collection. We adopt a L\'evy flight function to implement this mechanism because L\'evy flights explain exploration behaviors well in many social systems~\cite{riascos_random_2021, humphries_optimal_2014}.

In the third model, the next collection is chosen with a probability proportional to $1/d^{\alpha_{i}}$, where $d$ is the cosine distance between two collections' embedding vectors and $\alpha_i$ represents wallet $i$'s preference of choosing collections similar to the previous collection. If $\alpha$ is positive, a wallet tends to choose similar collections. If $\alpha$ is negative, a wallet tends to choose collections distant from the most recent collection. We name $\alpha$ as ``visual affinity exponent''. 

The likelihood of a trajectory $T_i$ is calculated from a prior distribution for a given $\alpha$. It is the multiplication of probabilities for distances between collections, based on a prior distribution $f(d|\alpha)$. In our previous example $\{d(C_1, C_2), d(C_2, C_3)\}$, its likelihood is $f(d(C_1, C_2)|\alpha) \times f(d(C_2, C_3)|\alpha)$ where $f(d|\alpha)$ is defined as: 

\begin{align}
\label{alpha}
\begin{split}
    f(d|\alpha) &= \frac{1}{C} \times \frac{f_{null}(d)}{d^\alpha}, \\
    C &= \sum_{d} \frac{f_{null}(d)}{d^\alpha} \\
\end{split}
\end{align}

\begin{figure}[h]
    \centering
    \includegraphics[width=0.6\textwidth]{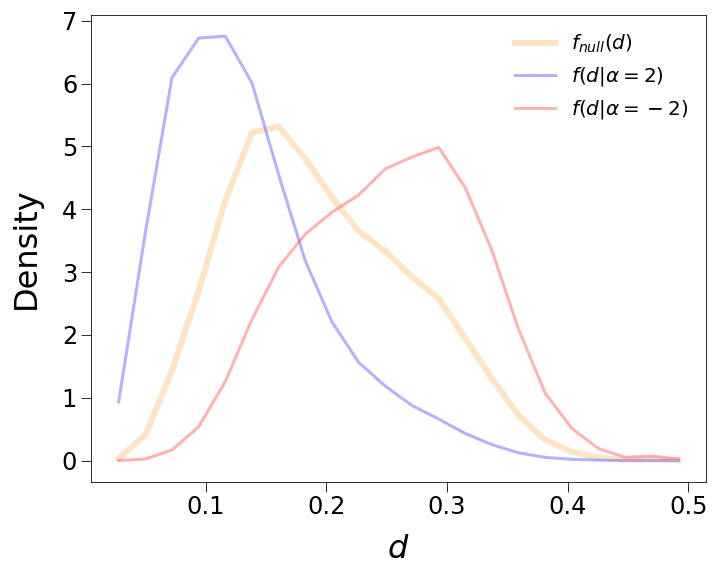}
    \caption{Two exemplar prior distributions and the null distribution. The null distribution represents the topology of the embedding space. If a wallet randomly selects the next collection for purchase, then the probability to choose a collection follows the null distribution. $f(d|\alpha=2)$ and $f(d|\alpha=-2)$ are prior distributions for the maximum likelihood method. If the visual affinity exponent $\alpha$ of a wallet is positive (blue curve), the wallet prefers to select similar collections. If $\alpha$ is negative (red curve), a wallet tends to choose different collections compared with the previous one.}
    \label{figure2}
\end{figure}

$f_{null}(d)$ and two exemplar $f(d|\alpha)$ distributions are shown in Figure~\ref{figure2}. The $f(d|\alpha=2)$ is shifted to the left compared to the null distribution, which means that wallets following this distribution prefer to purchase NFTs in collections similar to the previous collections. Note that $f_{null}(d)$ is identical to $f(d|\alpha=0)$. 

For each wallet, we estimate $\alpha$ maximizing the likelihood of obtaining its purchase trajectory and set it as the wallet's visual affinity exponent. Unfortunately, we can't calculate $\alpha$ values analytically because the collections are already distributed on the embedding space and their distributions are not represented with general functions. To make our estimation feasible, we change values from -6 to 6 with a step size of 0.1 and determine the value of the highest likelihood as $\alpha_i$ for a wallet $i$. 

\begin{align}
\alpha_i &= \argmax_\alpha \sum_{d \in T_i} \log{f(d | \alpha)}
\end{align}

\section{Results}
\label{Results}

\subsection{Model comparisons}

We generated 100 synthetic purchase trajectories for each wallet and compared them with its real purchase trajectory by calculating a distance error $DE$, which is the normalized average difference between the total distance of the real purchase trajectory and the total distance of a synthetic purchase trajectory. This value for a wallet $i$ is represented as 

\begin{align}
DE_{i} = \frac{1}{|T_i|-1} \frac{1}{100} \sum_{s=1}^{100} (\sum_{j=2}^{|T_i|} ({d(C_{i,j}, C_{i,j-1}) - d(C^s_{i,j}, C^s_{i,j-1}))}),
\end{align}

where $|T_i|$ is the length of $i$'s purchase trajectory, $j$ is the order of purchase in the trajectory, $s$ is the index of a synthetic trajectory, and $C^{s}_{i,j}$ is the $j$th collection in the $s$th synthetic trajectory.

\begin{figure}[h]
    \centering
    \includegraphics[width=0.8\textwidth]{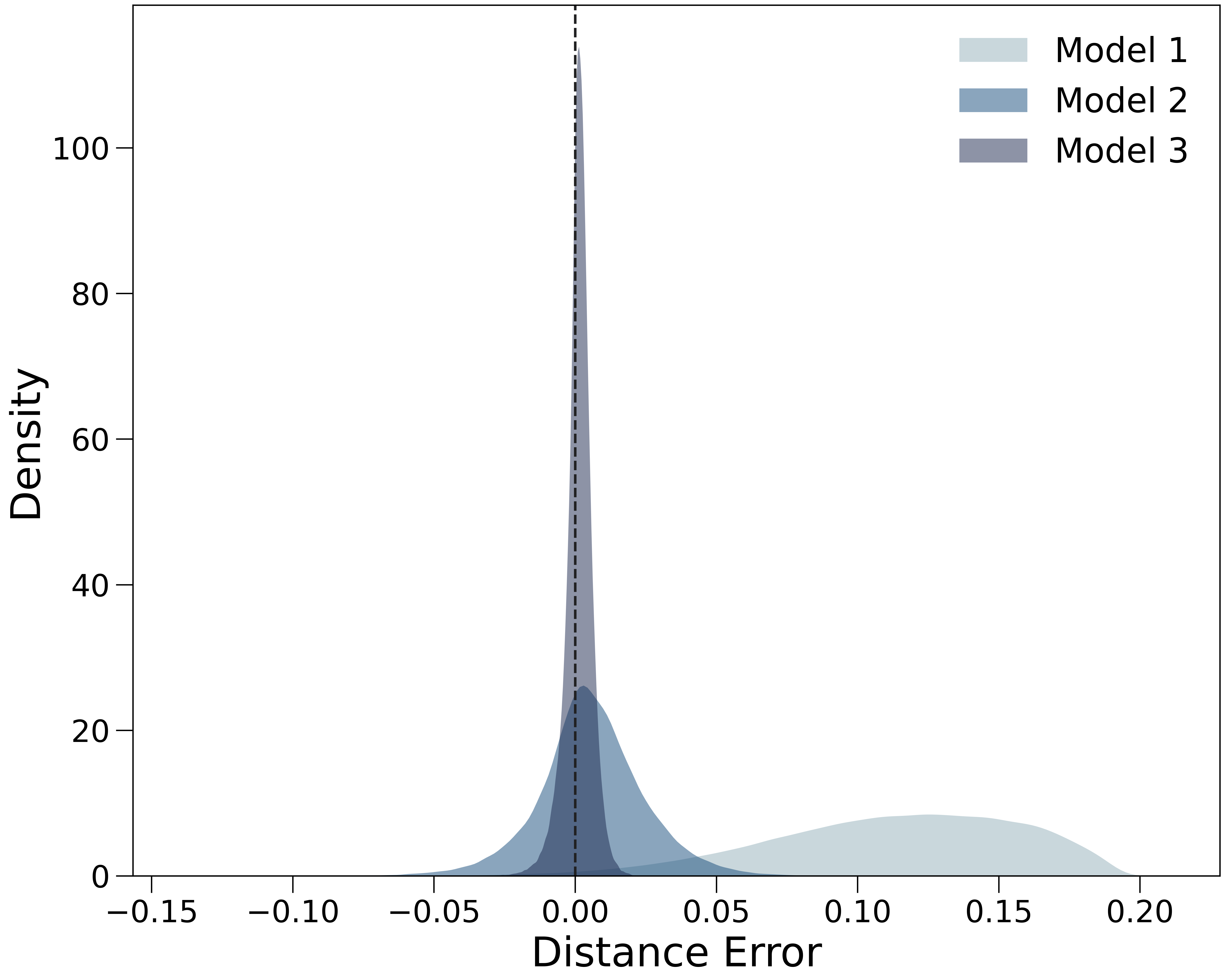}
    \caption{The distributions for distance errors of Model 1 (completely random), Model 2 (explore with a probability $p$ and choose next collections randomly), and Model 3 (explore with a probability $p$ and choose next collections by following a l\'evy flight function $d^{-\alpha}$). The closer a distribution is to the dashed line ($DE$=0), the better its associated model performs.}
    \label{model_comparision}
\end{figure}

Figure~\ref{model_comparision} shows how $DE$ values are distributed by model. Model 1, which creates random trajectories, has the highest error. Adding the exploration probability $p$ increases model performance significantly (Model 2). It suggests that exploitation with a probability of $1-p$ shapes purchase trajectories to a large extent. In addition to $p$, Model 3 incorporates a l\'evy flight function with the visual affinity exponent $\alpha$, which quantifies the tendency of purchase an NFT in a collection similar to the previous collections. We confirmed that Model 3 performs best with both the exploration probability $p$ and the visual affinity exponent $\alpha$. In the next section, we examine detailed properties and behaviors of wallets in the NFT market, especially focusing on estimated $p$ and $\alpha$ values.

\subsection{Interpreting wallet behaviors with estimated parameters}

\begin{figure}[h]
    \centering
    \includegraphics[width=\textwidth]{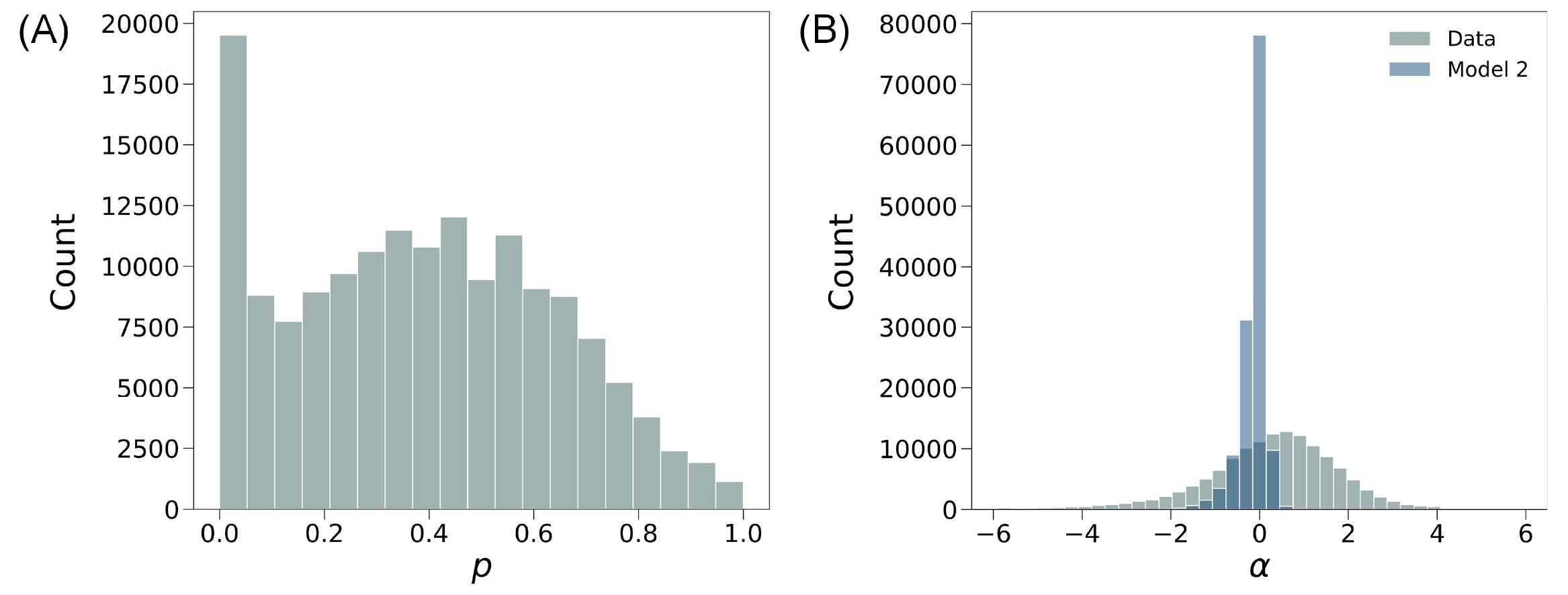}
    \caption{The distributions for the model parameters $p$ and $\alpha$. (A) $p$ is the exploration probability, which is equal to the proportion of inter-collection jumps. (B) $\alpha$ is the visual affinity exponent estimated from a L\'evy flight function. High $\alpha$ means the tendency to choose a similar collection for the next purchase. The distribution in gray shows estimated $\alpha$ values from 100 random sequences generated by Model 2.}
    \label{parameter_distribution}
\end{figure}

\subsubsection{Tendency to explore collections}

The estimated $p$ values suggest that there are mainly two groups of wallets (Figure~\ref{parameter_distribution}A). The first group consists of wallets exploiting only one collection ($p=0$). We would like to reiterate that the wallets which purchased more than 10 NFTs were analyzed, so this peak is not an artifact by low numbers of purchases. The other group comprises wallets exploring at least one collection ($p>0$). Estimated $p$ values are broadly distributed in this group. It indicates that the tendency to explore collections varies by wallet but not converges. Interestingly, there are more wallets of which $p$ value is lower than 0.5. The distribution implies that wallets in the NFT market tend to exploit a collection or explore relatively less. 

\subsubsection{Visual affinity to collections}

The visual affinity exponent $\alpha$ represents the extent to which a wallet favors new collections similar to the previous collections in its purchase trajectory. The distribution for $\alpha$ from the data is left-skewed (skewness=-0.6) and has a single peak at 0.7 with the median of 0.4 (Figure~\ref{parameter_distribution}B; The distribution colored in gray green). This pattern cannot be generated from random sequences. $\alpha$ values estimated from random sequences by Model 2 are distributed around 0 (Figure~\ref{parameter_distribution}B; The distribution colored in blue). The distribution implies that the NFT market consists of wallets having diverse preferences, but there are more wallets favoring collections similar to previous purchases (positive $\alpha$) compared to others exploring dissimilar collections (negative $\alpha$). On average, wallets behave in a conservative way in exploring collections while no physical risk and constraints exist in the embedding space.

\subsection{Application for a recommendation system}

\begin{figure}[h]
    \centering
    \includegraphics[width=0.75\textwidth]{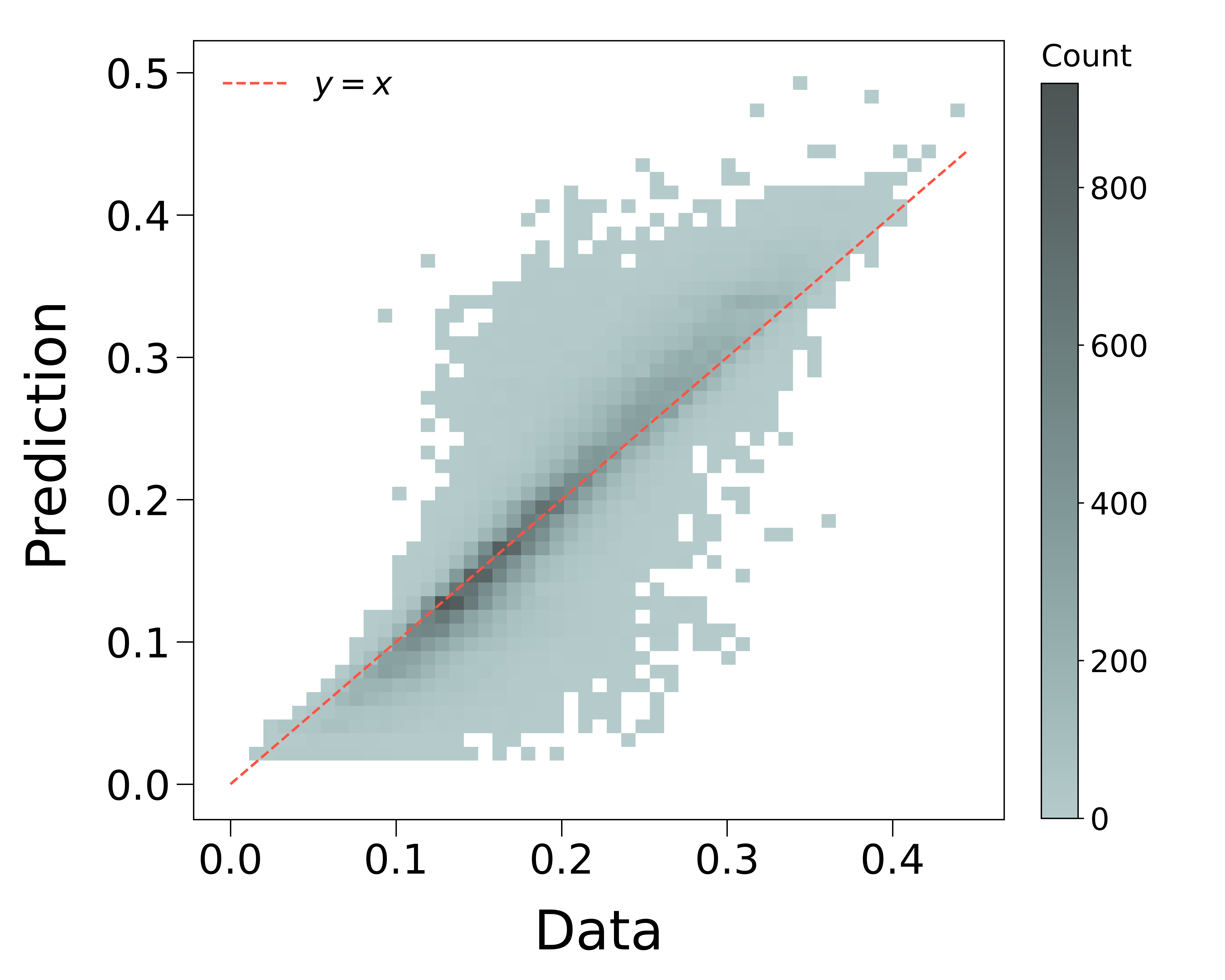}
    \caption{Performance of Model 3 as a recommendation system. Model 3 predicts the distance between the last collection and the second last collection in a trajectory. The predictions are close to the data as many incidences overlap with the $y=x$ line.}
    \label{res_recommend}
\end{figure}

We built a simple recommendation model predicting the last collection of a purchase trajectory with $p$ and $\alpha$ estimated from the collections in the trajectory except the last collection. For example, in a trajectory $[C_1, C_1, C_2, C_3]$, we exclude the last collection $C_3$ and estimate $p$ and $\alpha$ from $[C_1, C_1, C_2]$. We sampled the collection after $C_2$ 10 times ($X_1, ..., X_{10}$) from Model 3, calculated $d(C_2, X_1), ..., d(C_2, X_{10})$, and compared the average distance of $d(C_2, X_1), ..., d(C_2, X_{10})$ with the true value $d(C_2, C_3)$. We conducted this test only for wallets of which last and second to last collections are different. 
The average distances from predictions are highly similar to the actual data, resulting in the trend close to $y=x$ (Figure~\ref{res_recommend}). This result supports that $p$ and $\alpha$ are effective parameters describing wallet behaviors and can be used to build recommendation systems for NFT collections.

\section{Discussion}

Our results confirm that the exploration probability $p$ and the visual affinity exponent $\alpha$ characterize wallets' behaviors in selecting NFT collections well. The estimated parameters answer our research questions as below. 

For the first research question about how exploitation and exploration occur in the NFT market, we examined estimated $p$ values from the data. At the NFT collection level, exploitation and exploration correspond to the cases of $p=0$ and $p>0$, respectively. The proportion of exploitation is 10\%, which is not neglectable but indicates exploration is more popular in the NFT market. The broad $p$ distribution also shows that the extent of exploration varies by wallet (Figure~\ref{parameter_distribution}A). 

Our second research question is whether visual similarity affects exploration across NFT collections. We observed a large portion of the non-zero $\alpha$ values. It suggests visual similarity does matter in making exploration decisions (Figure~\ref{parameter_distribution}B). Specifically, as the mode and median of the $\alpha$ distribution are positive, we argue that wallets prefer similar collections when exploring the NFT market overall. 

The $\alpha$ distribution answers our third research question ``Do wallets in the NFT market explore similarly with agents in real-world physical systems?''. Literature about L\'evy flight reports that the exponent is between 1 and 3 in many physical systems~\cite{viswanathan2011physics}. Our result shows that there are indeed many wallets in this range but interestingly more wallets have an exponent below 1. Low exponents lead wallets to explore even less similar or dissimilar collections in the embedding space. 

This difference may be understood with the structure and characteristics of spaces and agents. Physical systems have limited reachable areas from a point, while our embedding space does not. Wallet owners can obtain any information about collections. Some wallets purchasing NFTs in collections distant in the embedding space can have negative $\alpha$. In addition, wallet owners may have diverse motivations in the NFT market. Some owners purchased NFTs for investment, while others did it for artistic collection purpose~\citep{yilmaz_what_2023}. In the latter case, wallets having a preference to specific visual components would show positive $\alpha$ values while the degree of the preference varies. 

However, we should acknowledge that these two parameters do not explain all variances in purchase decisions. First, considering all previous purchases is necessary to predict the next collection with high accuracy. Our current models only use the most recent purchase, so they basically rely on the Markov process. Due to its memoryless property, we don't leverage long-range correlations in purchase sequences. Advanced machine learning models designed for long sequences can be adopted to address this issue. Second, wallets' purchase decisions are affected by the popularity of a collection and its creator, the launching date of a collection, and the NFT price. These factors will add more dimensions in addition to visual similarity for understanding exploration and exploitation behaviors in the NFT market better.

\section{Conclusion}

We observed how visual similarity affects wallets' explorations in the NFT market by estimating two model parameters, exploration probability and visual affinity exponent. In general, wallets tend to explore NFT collections having similar visual features compared to the collections of their previous purchases. We also showed that our model parameters are useful for personalized recommendation systems. Our model can be improved with more factors of the NFT market and be extended to other digital asset markets.

\section*{Declaration of competing interest}
The authors declare that they have no known competing financial interests or personal relationships that could have appeared to influence the work reported in this paper. 

\section*{Acknowledgement}
This research was supported by the MSIT (Ministry of Science and ICT), Korea, under the ITRC (Information Technology Research Center) support program (IITP-2023-2018-0-01441) supervised by the IITP (Institute for Information \& Communications Technology Planning \& Evaluation). This work was supported by the National Research Foundation of Korea (NRF) grant funded by the Korea government (MSIT) (No. NRF-2021R1F1A1063030).

\section*{Data availability}
The data used in this manuscript is open source, and the sources are referenced in the manuscript. Codes will be available upon request.
\bibliographystyle{elsarticle-num} 
\bibliography{total.bib}

\begin{thebibliography}{10}
\expandafter\ifx\csname url\endcsname\relax
  \def\url#1{\texttt{#1}}\fi
\expandafter\ifx\csname urlprefix\endcsname\relax\def\urlprefix{URL }\fi
\expandafter\ifx\csname href\endcsname\relax
  \def\href#1#2{#2} \def\path#1{#1}\fi

\bibitem{nadini_mapping_2021}
M.~Nadini, L.~Alessandretti, F.~Di~Giacinto, M.~Martino, L.~M. Aiello, A.~Baronchelli, \href{https://www.nature.com/articles/s41598-021-00053-8}{Mapping the {NFT} revolution: market trends, trade networks, and visual features}, Scientific Reports 11~(1) (2021) 20902.
\newblock \href {https://doi.org/10.1038/s41598-021-00053-8} {\path{doi:10.1038/s41598-021-00053-8}}.
\newline\urlprefix\url{https://www.nature.com/articles/s41598-021-00053-8}

\bibitem{fridgen_pricing_2023}
G.~Fridgen, R.~Kräussl, O.~Papageorgiou, A.~Tugnetti, \href{https://papers.ssrn.com/abstract=4337173}{Pricing dynamics and herding behavior of {NFTs}}, preprint (2023).
\newblock \href {https://doi.org/10.2139/ssrn.4337173} {\path{doi:10.2139/ssrn.4337173}}.
\newline\urlprefix\url{https://papers.ssrn.com/abstract=4337173}

\bibitem{mekacher_heterogeneous_2022}
A.~Mekacher, A.~Bracci, M.~Nadini, M.~Martino, L.~Alessandretti, L.~M. Aiello, A.~Baronchelli, \href{https://www.nature.com/articles/s41598-022-17922-5}{Heterogeneous rarity patterns drive price dynamics in {NFT} collections}, Scientific Reports 12~(1) (2022) 13890.
\newblock \href {https://doi.org/10.1038/s41598-022-17922-5} {\path{doi:10.1038/s41598-022-17922-5}}.
\newline\urlprefix\url{https://www.nature.com/articles/s41598-022-17922-5}

\bibitem{la_cava_visually_2023}
L.~La~Cava, D.~Costa, A.~Tagarelli, \href{http://arxiv.org/abs/2303.17031}{Visually wired {NFTs}: Exploring the role of inspiration in non-fungible tokens}, preprint (2023).
\newblock \href {http://arxiv.org/abs/2303.17031 [physics]} {\path{arXiv:2303.17031 [physics]}}, \href {https://doi.org/10.48550/arXiv.2303.17031} {\path{doi:10.48550/arXiv.2303.17031}}.
\newline\urlprefix\url{http://arxiv.org/abs/2303.17031}

\bibitem{march_exploration_1991}
J.~G. March, \href{https://pubsonline.informs.org/doi/abs/10.1287/orsc.2.1.71}{Exploration and exploitation in organizational learning}, Organization Science 2~(1) (1991) 71--87.
\newblock \href {https://doi.org/10.1287/orsc.2.1.71} {\path{doi:10.1287/orsc.2.1.71}}.
\newline\urlprefix\url{https://pubsonline.informs.org/doi/abs/10.1287/orsc.2.1.71}

\bibitem{cohen_should_2007}
J.~D. Cohen, S.~M. McClure, A.~J. Yu, Should i stay or should i go? how the human brain manages the trade-off between exploitation and exploration, Philosophical Transactions of the Royal Society B: Biological Sciences 362~(1481) (2007) 933--942.

\bibitem{mehlhorn_unpacking_2015}
K.~Mehlhorn, B.~R. Newell, P.~M. Todd, M.~D. Lee, K.~Morgan, V.~A. Braithwaite, D.~Hausmann, K.~Fiedler, C.~Gonzalez, \href{http://doi.apa.org/getdoi.cfm?doi=10.1037/dec0000033}{Unpacking the exploration–exploitation tradeoff: A synthesis of human and animal literatures.}, Decision 2~(3) (2015) 191--215.
\newblock \href {https://doi.org/10.1037/dec0000033} {\path{doi:10.1037/dec0000033}}.
\newline\urlprefix\url{http://doi.apa.org/getdoi.cfm?doi=10.1037/dec0000033}

\bibitem{posen_chasing_2012}
H.~E. Posen, D.~A. Levinthal, \href{https://pubsonline.informs.org/doi/10.1287/mnsc.1110.1420}{Chasing a moving target: Exploitation and exploration in dynamic environments}, Management Science 58~(3) (2012) 587--601.
\newblock \href {https://doi.org/10.1287/mnsc.1110.1420} {\path{doi:10.1287/mnsc.1110.1420}}.
\newline\urlprefix\url{https://pubsonline.informs.org/doi/10.1287/mnsc.1110.1420}

\bibitem{phillips_rivals_2014}
N.~D. Phillips, R.~Hertwig, Y.~Kareev, J.~Avrahami, \href{https://www.sciencedirect.com/science/article/pii/S0010027714001164}{Rivals in the dark: How competition influences search in decisions under uncertainty}, Cognition 133~(1) (2014) 104--119.
\newblock \href {https://doi.org/10.1016/j.cognition.2014.06.006} {\path{doi:10.1016/j.cognition.2014.06.006}}.
\newline\urlprefix\url{https://www.sciencedirect.com/science/article/pii/S0010027714001164}

\bibitem{valone_group_1989}
T.~J. Valone, \href{https://www.jstor.org/stable/3565621?origin=crossref}{Group foraging, public information, and patch estimation}, Oikos 56~(3) (1989) 357.
\newblock \href {https://doi.org/10.2307/3565621} {\path{doi:10.2307/3565621}}.
\newline\urlprefix\url{https://www.jstor.org/stable/3565621?origin=crossref}

\bibitem{lee_exploration_2003}
J.~Lee, J.~Lee, H.~Lee, \href{https://pubsonline.informs.org/doi/abs/10.1287/mnsc.49.4.553.14417}{Exploration and {Exploitation} in the {Presence} of {Network} {Externalities}}, Management Science (Apr. 2003).
\newblock \href {https://doi.org/10.1287/mnsc.49.4.553.14417} {\path{doi:10.1287/mnsc.49.4.553.14417}}.
\newline\urlprefix\url{https://pubsonline.informs.org/doi/abs/10.1287/mnsc.49.4.553.14417}

\bibitem{schulz_structured_2019}
E.~Schulz, R.~Bhui, B.~C. Love, B.~Brier, M.~T. Todd, S.~J. Gershman, \href{https://www.pnas.org/doi/full/10.1073/pnas.1821028116}{Structured, uncertainty-driven exploration in real-world consumer choice}, Proceedings of the National Academy of Sciences 116~(28) (2019) 13903--13908.
\newblock \href {https://doi.org/10.1073/pnas.1821028116} {\path{doi:10.1073/pnas.1821028116}}.
\newline\urlprefix\url{https://www.pnas.org/doi/full/10.1073/pnas.1821028116}

\bibitem{reynolds_displaced_2007}
A.~M. Reynolds, A.~D. Smith, R.~Menzel, U.~Greggers, D.~R. Reynolds, J.~R. Riley, \href{https://onlinelibrary.wiley.com/doi/abs/10.1890/06-1916.1}{Displaced {Honey} {Bees} {Perform} {Optimal} {Scale}-{Free} {Search} {Flights}}, Ecology 88~(8) (2007) 1955--1961.
\newblock \href {https://doi.org/10.1890/06-1916.1} {\path{doi:10.1890/06-1916.1}}.
\newline\urlprefix\url{https://onlinelibrary.wiley.com/doi/abs/10.1890/06-1916.1}

\bibitem{reynolds_optimal_2008}
A.~M. Reynolds, \href{https://dx.doi.org/10.1209/0295-5075/82/20001}{Optimal random {Lévy}-loop searching: {New} insights into the searching behaviours of central-place foragers}, Europhysics Letters 82~(2) (2008) 20001.
\newblock \href {https://doi.org/10.1209/0295-5075/82/20001} {\path{doi:10.1209/0295-5075/82/20001}}.
\newline\urlprefix\url{https://dx.doi.org/10.1209/0295-5075/82/20001}

\bibitem{sims_encounter_2006}
D.~W. Sims, M.~J. Witt, A.~J. Richardson, E.~J. Southall, J.~D. Metcalfe, \href{https://royalsocietypublishing.org/doi/full/10.1098/rspb.2005.3444}{Encounter success of free-ranging marine predator movements across a dynamic prey landscape}, Proceedings of the Royal Society B: Biological Sciences 273~(1591) (2006) 1195--1201.
\newblock \href {https://doi.org/10.1098/rspb.2005.3444} {\path{doi:10.1098/rspb.2005.3444}}.
\newline\urlprefix\url{https://royalsocietypublishing.org/doi/full/10.1098/rspb.2005.3444}

\bibitem{focardi_adaptive_2009}
S.~Focardi, P.~Montanaro, E.~Pecchioli, \href{https://journals.plos.org/plosone/article?id=10.1371/journal.pone.0006587}{Adaptive {Lévy} {Walks} in {Foraging} {Fallow} {Deer}}, PLOS ONE 4~(8) (2009) e6587.
\newblock \href {https://doi.org/10.1371/journal.pone.0006587} {\path{doi:10.1371/journal.pone.0006587}}.
\newline\urlprefix\url{https://journals.plos.org/plosone/article?id=10.1371/journal.pone.0006587}

\bibitem{dai_short-duration_2007}
X.~Dai, G.~Shannon, R.~Slotow, B.~Page, K.~J. Duffy, \href{https://doi.org/10.1644/06-MAMM-A-035R1.1}{Short-{Duration} {Daytime} {Movements} of a {Cow} {Herd} of {African} {Elephants}}, Journal of Mammalogy 88~(1) (2007) 151--157.
\newblock \href {https://doi.org/10.1644/06-MAMM-A-035R1.1} {\path{doi:10.1644/06-MAMM-A-035R1.1}}.
\newline\urlprefix\url{https://doi.org/10.1644/06-MAMM-A-035R1.1}

\bibitem{brockmann_scaling_2006}
D.~Brockmann, L.~Hufnagel, T.~Geisel, \href{https://www.nature.com/articles/nature04292}{The scaling laws of human travel}, Nature 439~(7075) (2006) 462--465.
\newblock \href {https://doi.org/10.1038/nature04292} {\path{doi:10.1038/nature04292}}.
\newline\urlprefix\url{https://www.nature.com/articles/nature04292}

\bibitem{brown_levy_2007}
C.~T. Brown, L.~S. Liebovitch, R.~Glendon, \href{https://doi.org/10.1007/s10745-006-9083-4}{Lévy {Flights} in {Dobe} {Ju}/’hoansi {Foraging} {Patterns}}, Human Ecology 35~(1) (2007) 129--138.
\newblock \href {https://doi.org/10.1007/s10745-006-9083-4} {\path{doi:10.1007/s10745-006-9083-4}}.
\newline\urlprefix\url{https://doi.org/10.1007/s10745-006-9083-4}

\bibitem{gonzalez_understanding_2008}
M.~C. González, C.~A. Hidalgo, A.-L. Barabási, \href{https://www.nature.com/articles/nature06958}{Understanding individual human mobility patterns}, Nature 453~(7196) (2008) 779--782.
\newblock \href {https://doi.org/10.1038/nature06958} {\path{doi:10.1038/nature06958}}.
\newline\urlprefix\url{https://www.nature.com/articles/nature06958}

\bibitem{garg_efficient_2021}
K.~Garg, C.~T. Kello, \href{https://www.nature.com/articles/s41598-021-84542-w}{Efficient {Lévy} walks in virtual human foraging}, Scientific Reports 11~(1) (2021) 5242.
\newblock \href {https://doi.org/10.1038/s41598-021-84542-w} {\path{doi:10.1038/s41598-021-84542-w}}.
\newline\urlprefix\url{https://www.nature.com/articles/s41598-021-84542-w}

\bibitem{mandelbrot1982fractal}
B.~B. Mandelbrot, B.~B. Mandelbrot, The fractal geometry of nature, Vol.~1, New York: WH freeman, 1982.

\bibitem{zaburdaev_levy_2015}
V.~Zaburdaev, S.~Denisov, J.~Klafter, \href{https://link.aps.org/doi/10.1103/RevModPhys.87.483}{L{\textbackslash}'evy walks}, Reviews of Modern Physics 87~(2) (2015) 483--530.
\newblock \href {https://doi.org/10.1103/RevModPhys.87.483} {\path{doi:10.1103/RevModPhys.87.483}}.
\newline\urlprefix\url{https://link.aps.org/doi/10.1103/RevModPhys.87.483}

\bibitem{gleadell_phillips_2018}
\href{https://news.artnet.com/market/phillips-london-auction-march-2018-1240292}{Phillips {Steps} {It} {Up} {With} a {Triumphant} \$135 {Million} {Auction} in {London}, the {House}'s {Best} {Ever}}, (accessed 15 Jan 2024) (2018).
\newline\urlprefix\url{https://news.artnet.com/market/phillips-london-auction-march-2018-1240292}

\bibitem{opensea_top_nodate}
OpenSea, \href{https://opensea.io/rankings}{Top {NFTs}}, (accessed 15 Jan 2024) (2023).
\newline\urlprefix\url{https://opensea.io/rankings}

\bibitem{noauthor_blockchain-etlethereum-etl_nodate}
{Blockchain ETL}, \href{https://github.com/blockchain-etl/ethereum-etl}{blockchain-etl/ethereum-etl: {Python} scripts for {ETL} (extract, transform and load) jobs for {Ethereum} blocks, transactions, {ERC20} / {ERC721} tokens, transfers, receipts, logs, contracts, internal transactions. {Data} is available in {Google} {BigQuery} https://goo.gl/{oY5BCQ}}, (accessed 15 Jan 2024) (2019).
\newline\urlprefix\url{https://github.com/blockchain-etl/ethereum-etl}

\bibitem{he_deep_2016}
K.~He, X.~Zhang, S.~Ren, J.~Sun, Deep residual learning for image recognition, in: Proceedings of the IEEE conference on computer vision and pattern recognition, 2016, pp. 770--778.

\bibitem{riascos_random_2021}
A.~P. Riascos, J.~L. Mateos, \href{https://doi.org/10.1093/comnet/cnab032}{Random walks on weighted networks: a survey of local and non-local dynamics}, Journal of Complex Networks 9~(5) (2021) cnab032.
\newblock \href {https://doi.org/10.1093/comnet/cnab032} {\path{doi:10.1093/comnet/cnab032}}.
\newline\urlprefix\url{https://doi.org/10.1093/comnet/cnab032}

\bibitem{humphries_optimal_2014}
N.~E. Humphries, D.~W. Sims, \href{https://www.sciencedirect.com/science/article/pii/S0022519314003051}{Optimal foraging strategies: {Lévy} walks balance searching and patch exploitation under a very broad range of conditions}, Journal of Theoretical Biology 358 (2014) 179--193.
\newblock \href {https://doi.org/10.1016/j.jtbi.2014.05.032} {\path{doi:10.1016/j.jtbi.2014.05.032}}.
\newline\urlprefix\url{https://www.sciencedirect.com/science/article/pii/S0022519314003051}

\bibitem{viswanathan2011physics}
G.~M. Viswanathan, M.~G. Da~Luz, E.~P. Raposo, H.~E. Stanley, The physics of foraging: an introduction to random searches and biological encounters, Cambridge University Press, 2011.

\bibitem{yilmaz_what_2023}
T.~Yilmaz, S.~Sagfossen, C.~Velasco, \href{https://www.sciencedirect.com/science/article/pii/S0148296323004149}{What makes {NFTs} valuable to consumers? perceived value drivers associated with {NFTs} liking, purchasing, and holding}, Journal of Business Research 165 (2023) 114056.
\newblock \href {https://doi.org/10.1016/j.jbusres.2023.114056} {\path{doi:10.1016/j.jbusres.2023.114056}}.
\newline\urlprefix\url{https://www.sciencedirect.com/science/article/pii/S0148296323004149}

\end{thebibliography}

\pagebreak

\begin{center}
\textbf{\large{Supplementary Information}}
\end{center}

\setcounter{section}{0}
\setcounter{page}{1}

The following table includes 198 collections analyzed in the paper. These collections are selected from the OpenSea top 200 volume and trading counts lists on July 31, 2023. 

\begin{longtable}{l}
\toprule
Collections \\
\midrule
\endfirsthead
\toprule
Collections \\
\midrule
\endhead
\midrule
\multicolumn{1}{r}{Continued on next page} \\
\midrule
\endfoot
\bottomrule
\endlastfoot
"MOAR" by Joan Cornella \\
0N1 Force \\
0xApes Trilogy \\
10KTF \\
3Landers \\
8liens NFT \\
AINightbirds \\
Acrocalypse \\
Adam Bomb Squad \\
Akutars \\
Animetas \\
Ape Gang \\
Ape Kids Club (AKC) \\
Ape Reunion \\
Art Gobblers \\
Azuki \\
BEANZ Official \\
Bears Deluxe \\
Boki \\
Bored Ape Kennel Club \\
Bored Ape Yacht Club \\
Boss Beauties \\
C-01 Official Collection \\
CHIBI DINOS \\
CLONE X - X TAKASHI MURAKAMI \\
Capsule House \\
CatBlox Genesis Collection \\
Chain Runners \\
Cheebs NFT \\
Chimpers \\
Chungos \\
Cool Cats NFT \\
Cool Monkes Genesis \\
Cool Pets NFT \\
Coolman's Universe \\
Corntown wtf \\
Cosmic Cats Official \\
Cosmodinos Omega \\
Creature World \\
Creepz by OVERLORD \\
CrypToadz by GREMPLIN \\
Crypto Bull Society \\
Crypto.Chicks \\
CryptoBatz by Ozzy Osbourne \\
CryptoDickbutts \\
CryptoMories \\
CryptoNinja Partners \\
CryptoPunks \\
CryptoSimeji Official \\
CryptoSkulls \\
CyberKongz \\
CyberKongz VX \\
DEGEN TOONZ \\
DeGods \\
Deadfellaz \\
DentedFeelsNFT \\
Desperate ApeWives \\
DigiDaigaku Genesis \\
Dippies \\
Doodles \\
DopeApeClub \\
DourDarcels \\
EightBit Me \\
Elftown.wtf \\
FLUF World \\
FameLadySquad \\
Fang Gang \\
Fat Rat Mafia \\
Feline Fiendz NFT \\
FishyFam \\
Forgotten Runes Wizards Cult \\
FoxFam \\
Frenly Pandas \\
GOATz \\
GalacticApes \\
Galaxy Fight Club \\
Galaxy-Eggs \\
Genuine Undead \\
Girlies NFT \\
God Hates NFTees \\
Gold Hunt Game | GoldHunters \\
Goopdoods by Goopdude - Official Collection \\
HAPE PRIME \\
HYPEBEARSCLUB.OFFICIAL \\
Hashmasks \\
I Like You, You're Weird \\
IO: Imaginary Ones \\
Impostors Genesis Aliens \\
Inhabitants: Generative Identities \\
Invisible Friends \\
Isekai Meta \\
JUNGLE FREAKS GENESIS \\
KILLABEARS \\
KIWAMI Genesis \\
KPR \\
Kaiju Kingz \\
Kanpai Pandas \\
Karafuru \\
Killer GF \\
Kitaro World Official \\
Koala Intelligence Agency \\
Lazy Lions \\
Lil Pudgys \\
Lil' Heroes by Edgar Plans \\
Little Lemon Friends \\
Lives of Asuna \\
Lonely Alien Space Club \\
LonelyPop \\
Los Muertos World \\
Loser Club Official \\
MechMindsAI \\
Meebits \\
MekaVerse \\
Milady Maker \\
MoonCats \\
Moonbirds \\
Moonrunners Official \\
Murakami.Flowers Official \\
Muri by Fabrik \\
Mutant Ape Yacht Club \\
MutantCats \\
Nakamigos \\
NotOkayBears \\
Nyolings \\
OFFICIAL WAGMI ARMY \\
Okay Bears Yacht Club \\
OnChainMonkey \\
Outlaws \\
PEACEFUL GROUPIES \\
PXN: Ghost Division \\
PhantaBear \\
Pop Art Cats by Matt Chessco \\
Prime Ape Planet PAP \\
Project 0xD38 \\
Psychedelics Anonymous Genesis \\
Pudgy Penguins \\
Quirkies Originals \\
RENGA \\
Rare Apepes \\
Rare Land NFT \\
Redacted Remilio Babies \\
Robotos \\
Sappy Seals \\
Savage Nation \\
Seizon \\
Shellz Orb \\
Shinsekaicorp \\
SmallBrosNFT \\
Sneaky Vampire Syndicate \\
Society of Degenerate Apes (SODA) \\
Space Riders NFT \\
Starcatchers \\
SupDucks \\
SuperNormal \\
Swampverse \\
THE SHIBOSHIS \\
The Alien Boy \\
The Art of Seasons \\
The Doge Pound \\
The Heart Project \\
The Humanoids \\
The Long Lost \\
The Plague NFT \\
The Potatoz \\
The Sevens - Genesis \\
The Superlative Secret Society \\
The Weirdo Ghost Gang \\
The Wicked Craniums \\
Thingdoms \\
Trippy Toadz NFT \\
VOX Collectibles: Town Star \\
Valhalla \\
VeeFriends \\
WZRDS \\
We All Survived Death \\
We Are All Going to Die \\
Wicked Ape Bone Club \\
Winter Bears \\
WonderPals \\
World of Women \\
World of Women Galaxy \\
Wrapped Cryptopunks \\
YOLO HOLIDAY \\
Yaypegs \\
ZombieClub Token \\
a KID called BEAST \\
alien frens \\
feetpix.wtf \\
goblintown.wtf \\
ill poop it nft \\
inBetweeners by GianPiero \\
mfers \\
pablos.lol \\
rektguy \\
the littles NFT \\
tiny dinos (eth) \\
tubby cats \\
y00ts Yacht Club \\
\end{longtable}

\end{document}